\documentstyle[prl,multicol,aps,graphicx]{revtex}
\begin{document}
\draft

\title{Signal velocity, causality, and quantum noise in superluminal 
light pulse propagation}

\author{A. Kuzmich, A. Dogariu, and L. J. Wang\,* }
\address{NEC Research Institute, 4 Independence Way, Princeton, NJ 
08540, USA\\}
\author{P. W. Milonni}
\address{Los Alamos National Laboratory, Los Alamos, NM 87545, USA\\}
\author{R. Y. Chiao}
\address{Physics Department, University of California, Berkeley, CA 
94720, USA\\}
\date{submitted to Phys. Rev. Lett.)\\
(November 3, 2000}
\maketitle

\begin{abstract}

We consider pulse propagation in a linear anomalously dispersive medium where 
the group velocity exceeds the speed of light in vacuum ($c$) or even 
becomes negative. A signal velocity is defined operationally based on
the optical signal-to-noise ratio, and is computed for cases appropriate
to the recent experiment where such a negative group velocity was observed.
It is found that quantum fluctuations limit the signal 
velocity to values less than $c$. 
\end{abstract}
\begin{multicols}{2}
\narrowtext

It is well known that the group velocity $v_g$ of a light pulse can exceed 
$c$ in an anomalously dispersive medium. If there is no appreciable 
absorption or amplication, a sufficiently smooth initial pulse envelope 
$E(t)$ becomes simply $e^{i\phi}E(t-L/v_g)$ after the propagation distance $L$, 
where $\phi$ is a (real) phase. $E(t-L/v_g)$ in this case is the analytic 
continuation of $E(t-L/c)$ over the time increment $(1/c-1/v_g)L>0$ 
\cite{diener}. This analytic continuation means that information is 
transferred at velocity $c$, not $v_g$, so that there is no violation of 
causality implied by the superluminal group velocity. 
As discussed many years ago by Sommerfeld and Brillouin \cite{sb}, a group 
velocity greater than $c$ does not violate causality because it is not the
velocity of information transmission \cite{chiao}. They noted that the 
``frontal velocity," the velocity at which an infinitely sharp 
step-function-like disturbance of the light intensity propagates, can 
serve as a velocity of information transfer.

While a smoothly varying pulse is just an analytic continuation of
the input pulse $E(t-L/c)$, it is remarkable nonetheless that a very 
small leading edge enables one to predict the entire pulse. This small 
leading edge of the pulse can in principle extend infinitely far back in 
time.

These considerations are not immediately applicable in the laboratory. 
There is first of all the impossibility in principle of realizing the 
infinite bandwidth associated with a step-function ``front." But more 
subtle questions arise from the fact that a tiny leading edge of a smooth 
pulse determines the entire pulse. For one thing, it is not obvious how 
to define the ``arrival time" of the signal \cite{oppen}. In practice, 
one cannot extend the ``arrival time" to any time before the detection 
of the {\it first} photon. Furthermore, if the tiniest leading 
edge of a smooth ``superluminal" pulse determines the entire pulse, 
we must account for the effect that quantum fluctuations at 
the leading edge might have on the detection of the pulse \cite{ars},
\cite{segev}.

We suggest here an operational definition of the signal velocity and
apply it to the recently observed superluminal propagation of a
light pulse in a gain medium \cite{wang1}. This experiment showed 
not only that a superluminal group velocity is possible without any
significant pulse distortion, but also demonstrated 
that this can occur with no appreciable absorption or
amplification \cite{chu}. Previous considerations of quantum noise in 
this context focused on the motion of the peak of a wave packet, and
on the observability of the superluminal velocity of the peak at 
the one- or few-photon level \cite{ars},\cite{segev}. Here we consider
more generally the practical question of how a signal should be defined,
and reach the conclusion that quantum noise impedes the observation of 
the superluminal signal velocity, {\it regardless of the intensity of 
the input pulse.}

The experimental situation of interest is illustrated in Figure 1 
\cite{wang1}. A gas of atoms with a $\Lambda$-type transition
scheme is optically pumped into state $|1\rangle$. Two cw Raman pump
beams tuned off-resonance from the $|1\rangle\rightarrow|0\rangle$ 
transition with a slight frequency offset $2\Delta\nu$, and a 
pulsed probe beam acting on the $|0\rangle\rightarrow|2\rangle$ 
transition, propagate collinearly
through the cell. The common detuning $\Delta_0$ of the Raman and probe
fields from the excited state $|0\rangle$ is much larger than any of the
Rabi frequencies or decay rates involved, so that we can adiabatically
eliminate all off-diagonal density-matrix terms involving state
$|0\rangle$. Then we obtain the following expression for the linear 
susceptibility as a function of the probe detuning $\nu$ \cite{wang1},
\cite{wang2}:
\begin{equation}
\chi(\nu)={M\over\nu-\Delta\nu+i\gamma}+{M\over\nu+\Delta\nu+i\gamma} \ ,
\label{eq1}
\end{equation}
where $\gamma>0$ and $M>0$ is a two-photon matrix element whose detailed 
form and numerical value are not required for our present purposes. 
We note only that the dispersion relation (\ref{eq1}) satisfies the 
Kramers-Kronig relations and therefore that the pulse propagation 
described by it is causal. 

\begin{figure}
\hspace{.2in}
\includegraphics[scale=0.45]{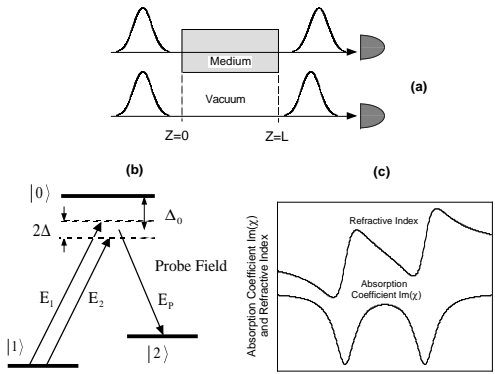}
\vspace{.2 in}
\caption
{(a) Schematic of the setup to create and observe transparent 
anomalous dispersion; (b) atomic transition scheme for double-peaked
Raman amplification; (c) refractive index and gain coefficient
as a function of probe beam frequency.}
\end{figure}

Consider now the detection of a signal corresponding to a light pulse
as indicated in Figure 1(a). We assign a time window $T$ centered about
a pre-arranged time $t_0$ at the detector and monitor the photocurrent
produced by the detector. We assume there is a background level
of irradiation that causes a constant average photocurrent $i_0$ when
no light pulse is received; there will be a nonvanishing $i_0$ whenever 
the medium exhibits gain.
We assume further that an increased 
photocurrent $i_1(t)$ is registered when a light pulse is received,
and assert that a {\it signal} has been received when the integrated 
photocurrent $\int dti_1(t)$ rises above the background level by a 
certain prescribed factor. The time at which this preset level of
confidence is reached is then {\it defined} to be the time of arrival
of this signal as recorded by an ideal detector.

The observable corresponding to this definition of the arrival time is 
\begin{equation}
\hat{S}(L,t) =C\int_{t_0-T/2}^tdt_1\hat{E}^{(-)}(L,t_1)
\hat{E}^{(+)}(L,t_1) \ ,
\label{eq2}
\end{equation}
where $\hat{E}^{(+)}(L,t_1)$ and $\hat{E}^{(-)}(L,t_1)$ are respectively
the positive- and negative-frequency parts of the electric field operator
at the exit point $(z=L)$ of the medium and $T/2$ is half the time window
assigned to the pulse, typically a few times the pulse width. $C$ is a 
constant to be specified later. The expectation value 
$\langle\hat{S}(L,t)\rangle$ 
is proportional to the number of photons that have arrived at the detector 
at the time $t$. If $\langle\hat{S}_1(L,t)\rangle$ and 
$\langle\hat{S}_0(L,t)\rangle$ are
respectively the expectation values of $\hat{S}(L,t)$ with and without 
an input pulse, then the photocurrent difference for an ideal
detector is $\langle\hat{S}_1(L,t)\rangle-\langle\hat{S}_0(L,t)\rangle$.
On the other hand, the second-order variance of the integrated photon number,
$\langle\Delta^2\hat{S}(L,t)\rangle$, gives the noise power due to quantum
fluctuations. Hence it is appropriate to define an optical signal-to-noise
ratio \cite{desurv}
\begin{equation}
SNR(L,t)={(\langle\hat{S}_1(L,t)\rangle-\langle\hat{S}_0(L,t)\rangle)^2
\over\langle\Delta^2\hat{S}(L,t)\rangle} \ .
\label{eq3}
\end{equation}
As discussed above, we define the arrival time $t_s$ of a signal as the 
time at which $SNR(L,t)$ reaches a prescribed threshold level.

The positive-frequency part of the electric field operator can be
written as
\begin{equation}
\hat{E}^{(+)}(z,t)=e^{-i\omega_o(t-z/c)}\int_{-\infty}^{\infty}dk
K(\omega)\hat{a}(\omega)e^{-i\omega(t-z/v_g)} \ ,
\label{eq4}
\end{equation}
where $\omega_o$ is the central frequency of the pulse, $K(\omega)
\propto\omega^{1/2}$, and $[\hat{a}(\omega),\hat{a}^{\dag}(\omega')]
=\delta(\omega-\omega')$. Eq. (\ref{eq4}) assumes plane-wave 
propagation in the $z$ direction and
that the group-velocity approximation is valid. 
$K(\omega)$ may for our purposes be taken to be constant since the frequency range
of the two gain lines are far smaller than the central frequency $\omega_{o}$.
It is then convenient to define the constant $C$ in Eq. (\ref{eq2}) to
be $1/(2\pi|K(\omega_{o})|^2)$.

In the experiment of interest the anomalously dispersive medium is a
phase-insensitive linear amplifier for which \cite{caves}
\begin{equation}
\hat{a}_{\rm out}(\omega)=g(\omega)\hat{a}_{\rm in}(\omega)+
\sqrt{|g(\omega)|^2-1}\hat{b}_{\rm in}^{\dag}(\omega) \ ,
\label{eq5}
\end{equation}
where $\hat{a}_{\rm in}$ and $\hat{a}_{\rm out}$ refer respectively 
to the input ($z=0$) and output ($z=L$) ports of the amplifier and
the operator $\hat{b}(\omega)$ is a bosonic operator ($[\hat{b}(\omega),
\hat{b}^{\dag}(\omega')]=\delta(\omega-\omega')$) that commutes with
all operators $\hat{a}_{\rm in}(\omega)$ and 
$\hat{a}_{\rm in}^{\dag}(\omega)$ and whose appearance in Eq. (\ref{eq5}) 
is required among other things to preserve the commutation relations 
for the field operators $\hat{a}_{\rm out}$ and $\hat{a}^{\dag}_{\rm out}$. 
$|g(\omega)|^2$ is the 
power gain factor characterizing the amplifier.

Now we derive a rather general expression for the optical
signal-to-noise ratio. Consider first the case of propagation over the
distance $L$ {\it in a vacuum} ($g(\omega)=1$). We assume that the 
initial state $|\psi\rangle$ of the field is a coherent state
such that $\hat{a}(\omega)|\psi\rangle=\alpha(\omega)|\psi\rangle$ 
for all $\omega$, where $\alpha(\omega)$ is a c-number. For such a
state we may write $\hat{E}^{(+)}(0,t)|\psi\rangle= \alpha(t)|\psi\rangle$,
where $\alpha(t)\equiv \pi^{-1/4}(N_p/\tau_p)^{1/2}
\exp(-(t-T_c)^2/2\tau_p^2)$, 
$N_p$ is the average number of photons in the initial pulse
of duration $\tau_p$, and $T_c$ is the time corresponding to the pulse 
peak. We obtain after a straightforward calculation the result
\begin{equation}
SNR(L,t)= \langle\hat{S}_1(0,t-L/c)\rangle_{\rm vac}= SNR(0,t-L/c) \ .
\label{eq6}
\end{equation}
This expresses the expected result that the pulse propagates at the 
velocity $c$ with no excess noise arising from propagation.

Next we treat the case of pulse propagation over the distance $L$
in the anomalously dispersive medium, using Eq. (\ref{eq5}) with
$g(\omega)\neq 1$ and assuming again an initially coherent field.
We obtain in this case 
\begin{equation}
\langle\hat{S}_1(L,t)\rangle-\langle\hat{S}_0(L,t)\rangle= 
|g(0)|^2\langle\hat{S}_1(0,t-L/v_g)\rangle_{\rm vac} 
\label{eq7}
\end{equation}
where $\langle\hat{S}_0(L,t)\rangle= (1/2\pi)\int_{T_c-T/2}^{t}dt_1
\int d\omega[|g(\omega)|^2-1]$ is the photon number in the absence 
of any pulse input to the medium and $T_c= t_0-L/v_g$.
The fact that $\langle\hat{S}_0(L,t)
\rangle>0$ is due to amplified spontaneous emission (ASE) \cite{desurv}; in
the experiment of interest the ASE is a spontaneous {\it Raman} process. 

Before proceeding further with the calculation of the optical 
signal-to-noise ratio, we note here that the gain factor 
\begin{equation}
|g(0)|^2= e^{4\pi M\gamma/(\Delta\nu^2+\gamma^2)\cdot L/\lambda} \ ,
\label{eq8}
\end{equation}
and that the effective signal $\langle\hat{S}_1(L,t)\rangle-
\langle\hat{S}_0(L,t)\rangle$ is proportional to the input
signal $\langle\hat{S}_1(0,t-L/v_g)\rangle_{\rm vac}$ with time
delay $L/v_g$ determined by the group velocity $v_g$. In the
anomalously dispersive medium $v_g= c/(n+\nu dn/d\nu)$ and
can be $>c$ or even negative, resulting in a time delay 
\begin{equation}
{L\over v_g}= {n_gL\over c}= \left[1-\nu_oM\cdot{\Delta\nu^2-\gamma^2
\over (\Delta\nu^2+\gamma^2)^2}\right]{L\over c} \ ,
\label{eq9}
\end{equation}
which is shorter than the time delay the pulse would experience upon
propagation through the same length in vacuum, or can become 
negative. In other words, the effective signal intensity can be 
reached sooner than in the case of propagation in vacuum.

In order to determine with confidence when a signal is received, however, one must
evaluate the SNR. Again using the commutation relations for the field 
operators, we obtain for the fluctuating noise background 
\begin{eqnarray}
\langle\Delta^2\hat{S}(L,t)\rangle &\equiv&
\langle\hat{S}^2(L,t)\rangle -\langle\hat{S}(L,t)\rangle^2 \nonumber \\
&=&|g(0)|^2\langle\hat{S}_1(0,t-L/v_g)
\rangle_{\rm vac} \nonumber \\
&+& 2{\rm Re}[|g(0)|^2\int_{t_0-T/2}^tdt_1
\int_{t_0-T/2}^tdt_2 \nonumber \\
&\times& \alpha^*(t_1-L/v_g)\alpha(t_2-L/v_g)F(t_1-t_2)] \nonumber \\
&+&\int_{t_0-T/2}^tdt_1\int_{t_0-T/2}^tdt_2|F(t_1-t_2)|^2 \ .
\label{eq10}
\end{eqnarray}
Here
\begin{equation}
F(t)=\int_{-\infty}^{\infty}d\omega[|g(\omega)|^2-1]e^{-i\omega t}
\label{eq11}
\end{equation}
is a two-time correlation function for the amplified spontaneous emission
noise. The three terms in Eq. (\ref{eq10}) can be attributed to amplified 
shot noise, beat noise, and ASE self-beat noise, respectively \cite{yam}.
The first two terms dominate in the presence of a strong input signal pulse,
while the last term dominates if the input signal is small and the gain
is large. In the case of a strong input signal and large gain, the second
term gives the largest contribution to the noise and scales almost
linearly with the signal strength $\langle\hat{S}\rangle$, the signal gain
$|g(0)|^2$, and the peak gain
$|g(\Delta\nu)|^2=\exp(4\pi ML/\gamma\lambda)$.
This large noise term effectively reduces the signal-to-noise ratio
of the advanced light pulse, causing the effective signal arrival
time to be retarded by a time delay that is far larger than the
pulse advancement.

Finally, using Eqs. (\ref{eq7}) and (\ref{eq10}), we compute 
$SNR^{(\rm out)}(L,t)$ for the propagation through 
the anomalously dispersive medium. In Figure 2 we plot the results
of such computations for $SNR^{(\rm out)}(L,t)$ as a function of
time on the output signal. For reference we also show $SNR$ for the
identical pulse propagating over the same length in vacuum. It is
evident from the results shown that the pulse propagating in vacuum
maintains a higher SNR. In other words, for the experiments of interest
here \cite{wang1},\cite{wang2}, {\it the actual arrival time of the
signal is delayed, even though the pulse itself is advanced 
compared with propagation over the same distance in vacuum.}

\begin{figure}
\hspace{.1in}
\includegraphics[scale=0.45]{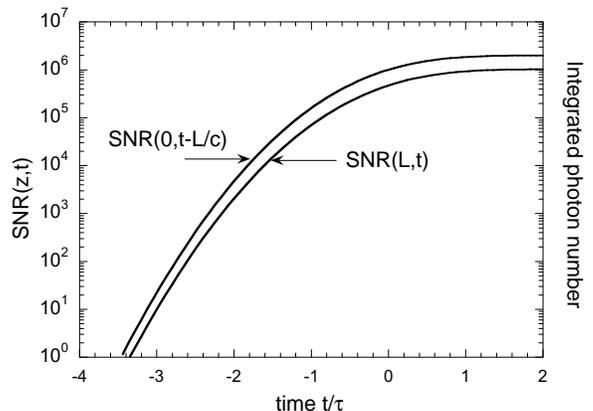}
\caption{Signal-to-noise ratios for light pulses propagating 
through the gain-assisted anomalous dispersion medium $SNR(L,t)$, 
and through the same distance in a vacuum $SNR(0,t-L/c)$.
Parameters used in the figure are adopted from the experiments 
reported in References [7] and [9].}
\end{figure} 

By requiring that at some time $t'$ the SNR of the output pulse be
equal to that of the input pulse [Eq. (\ref{eq6})] at a time $t$,
i.e.,
\begin{equation}
SNR^{({\rm out})}(L,t')=SNR^{({\rm in})}(0,t) \ ,
\label{eq12}
\end{equation}
we obtain a time difference $\delta t=t'-t$ that marks the propagation
time of the light signal, and $L/\delta t$ gives the signal velocity. In
Figure 3 we plot $\delta t$ as a function of gain for $(t-T_c)/\tau_p=
-1,-2,$ and $-3$. This corresponds to cases where the signal point is set
at 1,2, and 3 times the pulse width on the leading edge of the pulse. 
We also plot for reference the pulse advance $L/v_g$. It is evident that the 
retardation in the SNR far exceeds the pulse advancement. In other words,
the quantum noise added in the process of advancing a signal effectively
impedes the detection of the useful signal defined by the signal-to-noise 
ratio.

\begin{figure}
\includegraphics[scale=0.45]{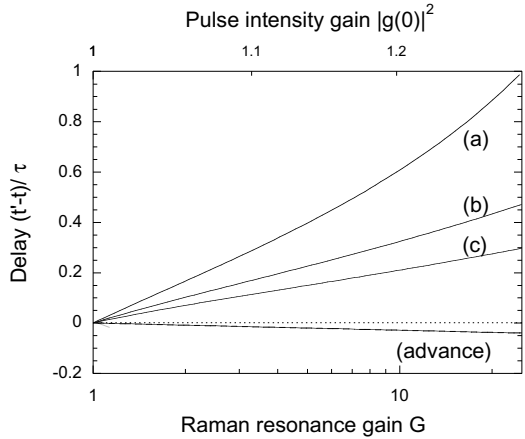}
\vspace*{.2 in}
\caption
{Delay $\delta t=t'-t$ due to reduced signal-to-noise ratio as a 
function of the gain coefficient. Curves (a), (b), and (c) are for 
$t/\tau_{p}=-1,-2,$ and $-3$, respectively.}
\end{figure}

In this letter we have presented what in our opinion is a
realistic definition, based on
photodetections, of the velocity of the signal carried by a light
pulse. We analyzed this signal velocity for 
the recently demonstrated superluminal light pulse propagation, and
found that while the pulse and the effective signal
are both advanced via propagation at a group velocity higher than 
$c$, or even negative, the signal
velocity defined here is still bounded by $c$. The physical mechanism 
that limits the signal velocity
is quantum fluctuation. Namely, because the transparent, 
anomalously dispersive medium is realized
using closely-placed gain lines, the various amplified quantum 
fluctuations introduce additional noise
that effectively reduces the SNR in the detection of 
the signals carried by the light pulse. This is related to the ``no 
cloning" theorem \cite{zurek},\cite{milmand},
which was attributed to the quantum fluctuations in an 
amplifier, and which is a direct consequence 
of the superposition principle in quantum theory.

Finally we note that it is perhaps possible to find other 
definitions of a ``signal" velocity for a light
pulse, different from that we presented here. But such a definition 
should in our opinion satisfy two basic criteria. First, it must be 
directly related to a known and practical way of detecting a signal.
Second, it should refer to the {\it fastest} practical way of communicating
information. While it may be hard to prove that any definition meets 
the second criterion, it can be hoped that the recent interest in 
quantum information theory might lead to a generally accepted
notion of the signal velocity of light.  

\acknowledgements{LJW wishes to thank R. A. Linke and J. A. Giordmaine 
for stimulating discussions.
\vspace{6 pt}\\
{ }* Email: Lwan@research.nj.nec.com}

\end{multicols}
\end{document}